\documentstyle{article}

\def\apj{ApJ}
\def\baas{BAAS}
\def\mnras{MNRAS}
\def\aap{A\&A}

\begin{document}

\def\bline{\rule[1.2mm]{3em}{0.1mm}}
\def\msun{M_\odot}
\def\zsun{Z_\odot}
\def\lsun{L_\odot}
\def\fun#1#2{\lower3.6pt\vbox{\baselineskip0pt\lineskip.9pt
  \ialign{$\mathsurround=0pt#1\hfil##\hfil$\crcr#2\crcr\sim\crcr}}}
\def\la{\mathrel{\mathpalette\fun <}}
\def\ga{\mathrel{\mathpalette\fun >}}
\def\eg{{\it e.g., }}
\def\etal{{\it et al. }}
\def\etalc{{\it et al., }}
\def\pc {{\rm pc}}
\def\mpc{{\rm Mpc}}
\def\kpc{{\rm kpc}}
\def\Mpc{{\rm Mpc}}

\def\he#1{\hbox{${}^{#1}{\rm He}$}}
\def\li#1{\hbox{${}^{#1}{\rm Li}$}}

\def\imfm{\xi_{\star}}
\def\mrem{m_{\rm rem}}
\def\avg#1{\langle #1 \rangle}
\def\sigbar{\avg{\sigma}}
\def\rbar{\hbox{$\avg{r}$}}
\def\omegam{\Omega_{\rm Macho}}
\def\omegab{\Omega_{\rm B}}
\def\omegalya{\Omega_{{\rm Ly}\alpha}}

\def\lya{Ly$\alpha$}

\def\pcite#1{(\cite{#1})}
\def\pref#1{(\ref{#1})}

\def\macho{{\sc macho}}
\def\newpage{\vfill\eject}
\def\vs{\vskip 0.2truein}
\def\gnu{\Gamma_\nu}
\def\fnu {{\cal F_\nu}}
\def\mass{m}
\def\lum{{\cal L}}
\def\imf{\Psi(\mass)}
\def\ilf{\Phi(M)}
\def\msun{M_\odot}
\def\zsun{Z_\odot}
\def\met{[M/H]}
\def\vi{(V-I)}
\def\mtot{M_{\rm tot}}
\def\mhalo{M_{\rm halo}}
\def\pp{\parshape 2 0.0truecm 16.25truecm 2truecm 14.25truecm}
\def\la{\mathrel{\mathpalette\fun <}}
\def\ga{\mathrel{\mathpalette\fun >}}
\def\fun#1#2{\lower3.6pt\vbox{\baselineskip0pt\lineskip.9pt
  \ialign{$\mathsurround=0pt#1\hfil##\hfil$\crcr#2\crcr\sim\crcr}}}
\def\ie{{\it i.e., }}
\def\eg{{\it e.g., }}
\def\etal{{\it et al. }}
\def\etalc{{\it et al., }}
\def\kpc{{\rm kpc}}
 \def\Mpc{{\rm Mpc}}
\def\mh{\mass_{\rm H}}
\def\mmax{\mass_{\rm u}}
\def\ml{\mass_{\rm l}}
\def\bc{f_{\rm cmpct}}
\def\br{f_{\rm rd}}
\def\kmsec{{\rm km/sec}}
\def\ibl{{\cal I}(b,l)}
\def\dmax{d_{\rm max}}
%
%
%




\centerline{\bf WHAT ARE MACHOS?}
\centerline{\bf LIMITS ON STELLAR OBJECTS}
\centerline{\bf AS THE DARK MATTER OF OUR HALO}

\bigskip
\renewcommand{\thefootnote}{\fnsymbol{footnote}}
\centerline{{{Katherine Freese}}, Brian Fields, and David Graff}
\footnote{\tt ktfreese@umich.edu; graff.25@osu.edu; bdfields@astro.uiuc.edu} 
\vskip 0.1in
\centerline{talk presented by {\bf Katherine Freese}}
\centerline{at the International Workshop on}
\centerline{Aspects of Dark Matter in Astro and Particle Physics,}
\centerline{Heidelberg, Germany, July 1998}

\vskip 0.1in
{\it\centerline{University of Michigan, Department of Physics}
\centerline{Ann Arbor, MI 48109-1120}}

\begin{abstract}
The nature of the Massive Compact Halo objects seen
in microlensing experiments and interpreted
as dark matter in the Halo
of our Galaxy remains a mystery.  Arguments are presented
that these events are probably not ordinary stellar or
substellar objects, i.e., they are probably not faint stars,
brown dwarfs, white dwarfs, or neutron stars.  On theoretical 
grounds one is then pushed to either exotic explanations
or a ``no-Macho" Halo (in which the Machos
reside elsewhere than in the Halo).  
Indeed a nonbaryonic component
in the Halo seems to be required.
\end{abstract}

\section{\bf Introduction}
\setcounter{footnote}{0}
\renewcommand{\thefootnote}{\arabic{footnote}}

The Halo of our Galaxy is made of as yet unidentified dark matter.
One of the outstanding questions in astrophysics is the nature
of this dark matter.  Microlensing experiments were
designed to look for $(10^{-7} - 1)\msun$ candidates, probably
baryonic.  These objects are called MACHOs, or Massive Compact
Halo Objects.  However, not only is the issue of ``What is the dark matter?"
still unresolved by the microlensing experiments,
but additional new puzzles have arisen.  I will argue for  my personal
conviction that 
\hfill\break
{\bf 1) Most of the dark matter in the Galactic Halo
must be nonbaryonic}, and \hfill\break
{\bf 2) The nature and origin of the Machos that have been interpreted
as being in the Halo are currently not at all understood.}

Until recently, stellar candidates for the dark matter in galaxies
were extremely popular.  However, in the last few years most
of these candidates have been either ruled out or shown to have
serious problems.  Using Hubble Space Telescope 
and parallax data (with some caveats
mentioned in the text), we showed that faint stars
and brown dwarfs contribute no more than 1\% of the mass
density of the Galaxy.  Recent microlensing events interpreted
as being in the Halo
have a best fit mass of $\sim 0.5\msun$, so that white dwarfs
have been taken seriously as dark matter candidates. 
However, stellar remnants including white dwarfs
and neutron stars are shown to be extremely problematic as
dark matter candidates.  It is a combination of mass budget issues
and chemical abundances that lead to the problems:
A significant fraction of the
baryons of the universe would have to be cycled through
the white dwarfs (or neutron stars) and their 
main sequence progenitors; however, in the process,
an overabundance of carbon and nitrogen is produced,
far in excess of what is observed both inside the Galaxy
and in the intergalactic medium.
Hence white dwarfs, brown dwarfs, faint stars,
and neutron stars are either ruled out or extremely problematic
as dark matter candidates.  Thus the puzzle remains,
What are the 14 MACHO events that have been interpreted as
being in the Halo of the Galaxy?  Are some of them actually
located elsewhere, such as in the LMC itself? These questions
are currently unanswered.

\subsection{\bf Microlensing Experiments}

The MACHO (Alcock \etal \cite{macho:1yr}, \cite{macho:2yr})
and EROS (Ansari \etal \cite{ansari}) experiments have
attempted to find the dark matter of our Galactic Halo by
monitoring millions of stars in the neighboring Large Magellenic
Cloud (LMC), which is approximately (45-60) 
kpc away; they have monitored stars in the Small Magellenic Cloud
(SMC) as well.  When a Macho crosses the line
of sight between a star in the LMC and us, 
the Macho's gravity magnifies the light of the background star.
The background star gets temporarily
brighter and then dims back down.  The Macho acts as a lens
for the background star.  The duration of the event scales as
$\Delta t \propto {\sqrt{m} \over v}$, where $m$ is the mass of the
Macho and $v$ is the velocity perpendicular to the line of sight.
Thus there is a degeneracy in the interpretation of the data
between $m$ and $v$.  To break the degeneracy, one has to assume
a galactic model, e.g., one has to assume that the lenses are in the Halo
of our Galaxy. 
The three events in the first year MACHO data had a typical
timescale of 40 days, which corresponds (with the above
assumption) to a best fit mass
for the Machos of $\sim 0.1\msun$.
With reanalysis and more data, four years of data yield 14 events 
of longer duration, 35-150 days (T. Axelrod
\cite{axelrod}; this is the Einstein diameter
crossing time).  Thus the new best fit mass is roughly 
$$m \sim 0.5 \msun \, . $$

From the experiments, one can estimate what fraction of the Halo
is made of Machos.  Using isothermal sphere models for the Galaxy with the
two year data, the Macho group estimated that 50\% (+30\%,-20\%)
of the Halo could be made of Machos.  However, this estimate
depends sensitively on the model used for the Galaxy. Gates, Gyuk,
and Turner \cite{ggt} ran millions of models and found that the number of 
models vs. Halo mass fraction peaks at Machos comprising (0-30)\% of the Halo,
with virtually no models compatible with a 100\% Macho Halo.

Hence there is evidence that a {\bf nonbaryonic} component to the
Halo of our Galaxy is required (see also the contribution
in this volume by Marc Moniez of the EROS group).

Microlensing experiments have ruled out a large class of 
possible baryonic dark matter components.  As described
in the contribution by Marc Moniez, substellar objects
in the mass range $10^{-7}\msun$ all the way up to $10^{-2}\msun$
are ruled out by the experiments.  In this talk
I will discuss the heavier possibilities in the range $10^{-2}\msun$
to few $\msun$. 

\section{\bf Baryonic Candidates}

In this talk I will concentrate on
baryonic candidates.   Hegyi and Olive \cite{ho} ruled
out large classes of baryonic candidates.  See also the work
of Carr \cite{carr}.
Until recently the most plausible remaining possibilities
for baryonic dark matter were \hfill\break

\noindent
--Red Dwarfs ($0.2 \msun >$ mass $> 0.09 \msun$).  These are stars just massive
enough to burn hydrogen; they shine due to fusion
taking place in the core of the star. Thus these are very faint
stars. \hfill\break

\noindent
--Brown Dwarfs (mass $< 0.09 \msun$).  These are sub-stellar
objects that cannot burn hydrogen.  They are too light to have
fusion take place in the interior. \hfill\break

\noindent
--White Dwarfs (mass $\sim 0.6 \msun$).  These are the end-products
of stellar evolution.

In this talk, I will present limits on red dwarfs 
(Graff and Freese 1996a), brown
dwarfs (Graff and Freese 1996b), and white dwarfs
(Graff, Laughlin, and Freese \cite{glf}; Fields, Freese, and Graff \cite{ffg};
Fields, Freese, and Graff \cite{ffg2})
as candidates for baryonic dark matter.

\section{Faint Stars and Brown Dwarfs}

Until recently, people noticed that the number of stellar
objects grows with decreasing stellar mass; hence there
was speculation that there might be a large number of faint
stars or brown dwarfs that are just too dim to have been seen.
However, as I will argue these candidates (modulo caveats below)
have now been ruled out as dark matter candidates.  Faint stars
and brown dwarfs constitute no more than a few percent of the
mass of our Galactic Halo.

\subsection{Faint Stars:}

First we used Hubble Space Telescope data (Bahcall, 
Flynn, Gould, and Kirhakos 1994) to limit the
mass density in red dwarfs to less than 1\% of the Halo
(Graff and Freese 1996a).   
The data of Bahcall et al (1994) from HST 
examined a small deep field and measured the relative
magnitudes of stars in the V and I bands.
We used the six stars that were seen with $1.7 < V-I < 3$ 
to limit the density of red dwarfs in the Halo.
First we obtained the distances to these stars, which are
shown in Figure 1.  One can see that the survey is sensitive
out to at least 10 kpc.  Note that the closest stars are likely
disk contaminants and not included in our final analysis.
We obtained estimates of the stellar
masses of these objects
from stellar models of Baraffe et al
(1996); the masses are in the range 0.0875$\msun$ - 0.2$\msun$.

For the 6 stars in the HST data with $1.7 < V-I < 3$,
we thus obtained a Halo red dwarf mass density.
We then compared this red dwarf mass density 
with virial estimates of the Halo
density to see what fraction is composed of red dwarfs.
We took a local Halo mass density of
$\rho_o \sim 9 \times 10^{-3} \msun/pc^3$.
Bahcall et al (1994) had made this comparison by assuming that the red
dwarfs had properties of stars
at the edge of the high metallicity main sequence;
these authors found that red dwarfs contribute less than 6\%
of the Halo density.
However, Halo red dwarfs are low metallicity objects,
and we were thus motivated to redo the analysis as
outlined above. A ground-based search for halo
red dwarfs by Boeshaar, Tyson, and Bernstein (1994) found a much
smaller number.  We felt that a careful reinterpretation
of the Bahcall et al (1994) data was in order.
Our result is that Red dwarfs with
$1.7 < V-I < 3$ (i.e., mass 0.0875 $< M/\msun < 0.2)$,
make up less than 1\% of the Halo; our best guess
is that they make up 0.14\% - 0.37\% of the mass of the halo.
Subsequent examination of the Hubble Deep
Field by Flynn, Gould, and Bahcall \cite{fgb} and work by
Mera, Chabrier, and Schaeffer \cite{mcs2} reiterated that
low-mass stars represent a negligible fraction of the
Halo dark matter.

\subsection{Brown Dwarfs:}

With these strong limits on the contribution
of faint stars to the Galactic Halo, we then obtained a Mass
Function of these same red dwarfs in order to be able
to extrapolate to the brown dwarf regime; in this way
we were able to limit the contribution of brown dwarfs as well.
We obtained the mass function from the following relation:
\begin{equation}
\label{derivemassfunc}
{\rm Mass Function} = {({\rm dM}_{\rm V} / {\rm dm})} \times  {\rm
Luminosity Function} \, .
\end{equation}
Here, the Mass Function (hereafter MF) is the number
density of stars with mass between $m$ and $m+dm$,
and the Luminosity Function (hereafter LF) is the number density of 
stars in a magnitude range $M \rightarrow M+dM$ (note that $M$
refers to magnitude while $m$ refers to mass).
The luminosity function is what is observed; we used parallax
data taken by the US Naval Observatory (Dahn et al 1995) who 
found 114 halo stars.
We went from this observed luminosity function to the desired
mass function via stellar models of $M_V(m)$ obtained
by Alexandre et al. \cite{models96}.

The parallax data (Dahn et al 1995) are shown in Figure 2. This is
an H-R diagram of nearby stars with measured parallax.  The filled circles
are high metallicity disk stars. The open circles are red
dwarfs which are known to be in the Halo because of their
low metallicities and high velocities.    It is these 114 Halo stars
that we used to get a mass function.  
We always took the most ``conservative" case, i.e., the
steepest MF towards low mass; this case would give the largest
number of brown dwarfs and low mass red dwarfs.
For this reason, we considered a number of metallicities
and used the lowest realistic value of $Z = 3 \times 10^{-4}$.
There is a potential complication
in that some of the stars in the survey
may actually be unresolved binaries.  If so, the observed
light is the sum of the light from two stars.  Then one 
may overestimate the mass of the star if one assumes 
the light is from a single star.  
We considered three
models for binaries.  The most extreme of these is that
all the stars are really in binaries, with equal masses
for the two stars in the binary system.  Then the luminosity
of each star is really half as big as if it had been a single
star, each star has a smaller mass, and one obtains
a steeper mass function towards low mass.  
This model is unphysical but simple, and we
used it to illustrate an extreme for
the largest number of stars at low mass that can
be obtained from this data set.
Figure 3 shows the mass functions that we obtained,
for the case of no binaries and the extreme case of 100\% binaries.
In these plots we  multiplied the vertical axis by $m^2$ for simplicity of
interpretation.   With this factor of $m^2$,
a mass function (MF) that is decreasing to the left
converges, an MF that is increasing to the left diverges, while
an MF that is flat diverges only logarithmically.  In figure 3a,
the case of no binaries, we can see that the MF $\times m^2$
decreases to the left (convergent); in Figure 4c, the case
of 100\% binaries, the MF $\times m^2$ is flat (diverges logarithmically).
Hence Figure 3 summarizes our results for the mass function for faint stars
heavier than $0.09\msun$. 

Now, in order to proceed with an extrapolation of this
red dwarf mass function past the hydrogen burning limit
into the red dwarf regime, we need a brief theoretical
interlude.  Star formation theory indicates that, as
one goes to lower masses, the MF rises no faster than
a power law.  The theories of Adams and Fatuzzo (1996),
Larson (1992), Zinnecker (1984), and Price
and Podsiadlowski (1995), while based on different
physical principles, all find this same upper limit.
Hence we looked for the power law describing the red
dwarf mass function at the lowest masses, and then
use this same power law to extrapolate into the brown
dwarf regime.  We took the mass function to scale as
\begin{equation}
\label{massfunc}
MF\propto m^{- \alpha} \, .
\end{equation}
Then the total mass in the Halo is
\begin{equation}
\label{mtot}
m_{tot} = \int_0^{0.09\msun} m \times MF \times dm \, . 
\end{equation}
If $\alpha > 2$, then the total mass diverges.  If $\alpha = 2$,
then the total mass diverges only logarithmically.  If $\alpha <2$,
then the total mass converges.  We found
\begin{equation}
\label{alpha}
\alpha \leq 2 \, , 
\end{equation}
for all models.  More specifically, for the extreme case
of 100\% binaries, we found $\alpha = 2$, i.e., each order of
magnitude of mass range contains an equal total mass.  Even for
a lower limit $\sim m_{moon}$, the total mass in brown dwarfs
is less than 3\% of the Halo mass.  For all other models,
including the case of no binaries, we find $\alpha <2$,
and brown dwarfs consitute less than a percent of the Halo mass.
Similar results were found by Mera, Chabrier, and Schaeffer \cite{mcs}.

Dalcanton \etal \cite{dal} found similar results by looking for
a reduction in apparent equivalent width of quasar emission lines;
such a reduction would be caused by compact objects such as brown dwarfs.

The two year MACHO microlensing data have also shown that,
for standard Halo models as well as a wide range of alternate models,
the timescales of the events are not compatible with a population
of stars lighter than 0.1$\msun$ (Gyuk, Evans, and Gates \cite{GEG}).

{\it Caveats:}
How might one avoid these conclusions?  First, star
formation theory might be completely wrong.
Alternatively,
there might be a spatially varying initial mass function so that
brown dwarfs exist only at large radii and not in our locality,
so that they were missed in the data (Kerins and Evans \cite{kerev}).

\subsection{Punchline:}
The basic result of this work is that the total
mass density of local Population II Red Dwarfs and
Brown Dwarfs makes up less than 1\% of the local mass
density of the Halo; in fact, these objects probably make up
less than $0.3\%$ of the Halo.

\section{Mass Budget Issues} 

This section (based on work by Fields, Freese, and Graff \cite{ffg})
is general to all Halo Machos, no matter what
kind of objects they are.

\subsection{Contribution of Machos to the Mass Density of the Universe:}
There is a potential problem in that too many baryons are
tied up in Machos and their progenitors (Fields, Freese, and Graff).
We begin by estimating the contribution of Machos to the mass density of the
universe:
Microlensing results (Alcock \etal 1997a) predict that the total mass
of Machos in the Galactic Halo out to 50 kpc is
\begin{equation}
\label{outto}
M_{\rm Macho} = (1.3 - 3.2) \times 10^{11} \msun \, . 
\end{equation}
Now one can obtain a ``Macho-to-light" ratio for the Halo by
dividing by the luminosity of the Milky Way (in the B-band),
\begin{equation}
\label{lummw}
L_{MW} \sim (1.3-2.5) \times 10^{10} L_\odot \, . 
\end{equation}
We obtain
\begin{equation}
\label{masstolight}
(M/L)_{\rm Macho} = (5.2-25)\msun/L_\odot \, . 
\end{equation}
From the ESO Slice Project
Redshift survey (Zucca \etal \cite{zuc}), 
the luminosity 
density of the Universe in the $B$ band is 
\begin{equation}
\label{phi}
{\cal L}_B = 1.9\times 10^{8} h \ L_\odot \ {\rm Mpc}^{-3}
\end{equation}
where the Hubble parameter 
$h=H_0/(100 \, {\rm km} \, {\rm sec}^{-1} \, \Mpc^{-1})$.  
If we assume that the $M/L$ which we defined for the Milky 
Way is typical of the Universe as a whole, 
then the universal mass density of 
Machos is 
\begin{equation}
\label{rho}
\rho_{\rm Macho} 
   = (M/L)_{\rm Macho} \, {\cal L}_B
   = (1-5) \times 10^{9} h \ \msun \, {\rm Mpc}^{-3} \, .
\end{equation}
The corresponding fraction of the critical density
$\rho_c \equiv 
3H_0^2/8 \pi G = 2.71 \times 10^{11} \, h^2 \, M_\odot \ \Mpc^{-3}$ is
\begin{equation}
\label{omega}
\Omega_{\rm Macho} \equiv \rho_{\rm Macho}/ \rho_c = (0.0036-0.017) \, h^{-1} \, .
\end{equation}

We will now proceed to compare our $\omegam$
derived in Eq.\ \pref{omega} with the baryonic density in the universe,
$\omegab$, as determined by primordial nucleosynthesis.
Recently, the status of Big Bang
nucleosynthesis has been the subject of
intense discussion, prompted both by observations of deuterium
in high-redshift quasar absorption systems, and also by
a more careful examination of consistency and uncertainties
in the theory.
To conservatively allow for the full range of possibilities,
we will therefore adopt
\begin{equation}
\label{omegab}
\omegab= (0.005-0.022) \ h^{-2} \, .
\end{equation}

We can see that
$\omegam$ and $\omegab$ are roughly comparable within this na\"{\i}ve 
calculation.  Thus, if the Galactic halo Macho
interpretation of the microlensing
results is correct,
Machos make up an important fraction of the baryonic matter 
of the Universe.  
Specifically, the central values in
eqs.\ (\ref{omega}) and (\ref{omegab}) give
\begin{equation}
\label{central}
\omegam/\omegab \sim 0.7 \, .
\end{equation}
However, the lower limit on this fraction is
considerably smaller and hence less restrictive.
Taking the lowest possible
value for $\omegam$ and the highest possible value for $\omegab$,
we see that
\begin{equation}
\label{comp}
{\omegam \over \omegab} \geq {1 \over 6} h \geq \frac{1}{12} \, .
\end{equation}

The only way to avoid these conclusions is to argue
that the luminosity density in eqn. (8) is dominated
by galaxies without Machos, so that the Milky Way is atypically
rich in Machos.  However, this is extremely unlikely,
because most of the light contributing to
the luminosity density ${\cal{L}}$  comes from galaxies similar to ours.
Even if Machos only exist in spiral galaxies (2/3 of the galaxies)
within one magnitude of the Milky Way, the value of $\Omega_{\rm Macho}$
is lowered by at most a factor of 0.17.

\subsection{Comparison with the Lyman-${\bf \alpha}$ Forest}

We can compare the Macho contribution to other components
of the baryonic matter of the universe.  In particular,
measurements of the Lyman-$\alpha$ (\lya) forest absorption from 
intervening gas in the lines of sight to high-redshift
QSOs indicate that many, if not most, of the baryons
of the universe were in this forest at redshifts $z>$2.
It is hard to reconcile the large baryonic abundance
estimated for the \lya\ forest with $\omegam$
obtained previously (Gates, Gyuk, Holder, \& Turner \cite{gght}).  
Although measurements of the \lya\
forest only obtain the neutral column density, careful
estimates of the ionizing radiation can be made to 
obtain rough values for the total baryonic matter,
i.e. the sum of the neutral and ionized components,
in the \lya\ forest.  For the sum of these
two components,
Weinberg et al. \pcite{wmhk} estimate
\begin{equation}
\label{lyman}
\Omega_{\rm Ly\alpha} \sim 0.02 h^{-3/2} \, .
\end{equation}
This number is at present uncertain.  For example, it
assumes an understanding of the UV background responsible
for ionizing the IGM, and accurate determination
of the quasar flux decrement due to the neutral
hydrogen absorbers.  Despite these uncertainties,
we will use Eq.\ ({\ref{lyman}}) below
and examine the implications of this estimate.

We can now require that the sum of the Macho energy density
plus the \lya\ baryonic energy density do not
add up to a value in excess of the baryonic density from
nucleosynthesis:
\begin{equation}
\label{insist}
\omegam(z) + \omegalya (z) \leq \omegab \, ;
\end{equation}
this expression holds for any epoch $z$.  
Unfortunately, the observations of Machos and 
\lya\ systems are available for different epochs.
Thus, to compare the two one must assume
that there has not been a tradeoff of gas into
Machos between the era of the Lyman systems
($z \sim 2-3$) and the observation of the Machos at $z=0$.
That is, we assume that the Machos were formed
before the \lya\ systems. 

Although Eq.\ (\ref{insist}) offers a potentially
strong constraint, in practice the uncertainties in both 
$\omegalya$ and in $\omegab$ make a quantitative
comparison difficult.
Nevertheless, we will tentatively use
the numbers indicated above.
We then have 
\begin{equation}
\label{onehalf}
(\omegam = 0.007-0.04) + (\omegalya = 0.06) \leq (\omegab = 0.02 - 0.09)
\ \ \ {\rm for} \, h=1/2 \, ,
\end{equation}
and 
\begin{equation}
\label{one}
(\omegam = 0.004 - 0.02) + (\omegalya = 0.02) \leq (\omegab = 0.005 - 0.02)
\ \ \ {\rm for} \, h=1 \, .
\end{equation}
These equations can be satisfied, but
only if one uses the most favorable extremes
in both $\omegam$ and $\omegab$, i.e., for
the lowest possible values for $\omegam$
and the highest possible values for $\omegab$.

Recent measurements of Kirkman and Tytler \pcite{kt}
of the ionized component of a Lyman limit system
at z=3.3816 towards QSO HS 1422+2309
estimate an even larger value for the mass density in 
hot and highly ionized gas in the intergalactic
medium: $\Omega_{\rm hot} \sim 10^{-2} h^{-1}$.
If this estimate is correct, then
Eq.\ (\ref{insist}) becomes even more
difficult to satisfy.

One way to avoid this mass budget problem would
be to argue that the \lya\ baryons later
became Machos.  Then it would be inappropriate to
add the \lya\ plus Macho contributions in
comparing with $\omegab$, since the Machos would
be just part of the \lya\ baryons.  However,
the only way to do this would be to make the Machos
at a redshift after the \lya\ measurements were
made.  Since these measurements extend down to about
$z\sim 2-3$, the Machos would have to be made
at $z<2$.  However, this would be difficult to maneuver.
A large, previously unknown population of
stellar remnants could not have formed after redshift 2; we would
see the light from the stars in galaxy counts (Charlot and Silk \cite{cs})
and in the Hubble Deep Field (Loeb \cite{loeb}).

Until now we have only considered the contribution to the baryonic
abundance from the Machos themselves. Note: see also the 
discussion by Fukugita, Hogan, and Peebles \cite{fhp}.  Below we will consider
the baryonic abundance of the progenitor stars as well, in the
case where the Machos are stellar remnants.  
When the progenitor baryons are added to the left hand side
of Eq.\ (\ref{insist}), this equation becomes harder to satisfy.
However, we wish to reiterate that measurements of $\Omega_{{\rm Ly} \alpha}$
are at present uncertain, so that it is possibly premature to
conclude that Machos are at odds with the amount of baryons in 
the Ly$\alpha$ forest.

\section{Machos as Stellar Remnants: White Dwarfs or Neutron Stars}

In the last section on the mass budget of Machos, we assumed merely
that they were baryonic compact objects.  In this section
(based on work by Fields, Freese, and Graff \cite{ffg}): 
we turn to the specific possibility that Machos are stellar remnants
white dwarfs, neutron stars, or black holes.  
The most complete microlensing data indicate a best fit mass
for the Machos of roughly 0.5$\msun$.  Hence there has
been particular interest in the possibility that these objects
are white dwarfs.  I will discuss problems and issues
with this interpretation: in particular I will discuss the
baryonic mass budget and the pollution due to white dwarf
progenitors.

\subsection{Mass Budget
Constraints from the Macho Progenitors:}

In general, white dwarfs, neutron stars, or black holes
all came from significantly heavier progenitors.  Hence, the excess mass
left over from the progenitors must be added to the calculation
of $\Omega_{\rm Macho}$; the excess mass then leads to stronger
constraints.
Previously we found that any baryonic Machos that are responsible
for the Halo microlensing events must constitute a significant
fraction of all the baryons in the universe.
Here we show that, if the Machos are white dwarfs or neutron
stars, their progenitors, while on the main sequence, 
are an even larger fraction of the total baryonic content of the universe.
The excess mass is then ejected in the form of gas when
the progenitors leave the main sequence and become stellar remnants.
This excess mass is quite problematic, as there is a lot
of it and it is chemically enriched beyond what is allowed by observations.

If all the Machos formed in a single burst (the burst model), then 
(for different choices of the initial mass function)
we can determine the additional contribution of the 
excess gas to the mass density of the universe.  Typically
we find the contribution of Macho progenitors to
the mass density of the universe to be
\begin{equation}
\label{omegaprog}
\Omega_{{\rm prog}} = 4 \Omega_{{\rm Macho}} = (0.016-0.08)h^{-1} \, . 
\end{equation}
(As an extreme minimum, we find an enhancement factor of 2 rather than 4).
From comparison with $\omegab$,
we can see that a very large fraction of the baryons of the
universe must be cycled through
the Machos and their progenitors.  
In fact, the central values of all the numbers now imply
\begin{equation}
\label{toomuch}
\Omega_{\rm prog} \sim 3 \Omega_B \, ,
\end{equation}
which is obviously unacceptable.  One is driven to the lowest
values of $\Omega_{rm Macho}$ and highest value of $\Omega_B$
to avoid this problem.

\subsection{Galactic Winds}
The white dwarf progenitor stars
return most of their mass
in their ejecta, i.e., planetary nebulae 
composed of processed material.
Both the mass and the composition of the material
are potential problems.  As we have emphasized, 
the cosmic Macho mass budget is a serious issue.
Here we see that it is significant
even when one considers only the Milky Way.
The amount of mass ejected by the progenitors is far in excess
of what can be accomodated by the Galaxy.  
Given the $M_{\rm Macho}$ of Eq.(5), 
a burst model requires the total mass of progenitors in the Galactic Halo
(out to 50 kpc) to have been at least twice the total mass in remnant
white dwarfs, i.e.,
$M_{{\rm prog}} \ge 2 M_{\rm Macho} = (2.4 - 5.8) \times 10^{11} \msun$.
The gas that is ejected by the Macho
progenitors is collisional and tends to fall into the Disk
of the Galaxy.
But the mass of the ejected gas $M_{\rm gas}
= M_{{\rm prog}} - M_{\rm Macho} \sim M_{\rm Macho}$
is at least as large as the mass ($\sim 10^{11} M_\odot$)
of the Disk and Spheroid of the
Milky Way combined.  Thus the gas
ejected by the Macho progenitors exceeds the mass of the Disk
and Spheroid. 
We see that the Galaxy's baryonic mass budget---including Machos---immediately 
demands that 
some of the ejecta be {\it removed} from the Galaxy.

This requirement for outflow is intensified when 
one considers the composition of the stellar ejecta.
It will be void of deuterium, and will include large amounts
of the nucleosynthesis products of $(1-8) \msun$ white
dwarf progenitors, notably:  helium, carbon, and nitrogen 
(and possibly s-process material).

A possible means of removing these excess baryons
is a Galactic wind.  Indeed, as pointed out
by Fields, Mathews, \& Schramm \pcite{fms}, 
such a wind may be a virtue, as hot gas containing
metals is ubiquitous in the universe, seen in
galaxy clusters and groups, and present as an ionized
intergalactic medium that dominates the observed
neutral \lya\ forest.  Thus, it seems mandatory
that many galaxies do manage to shed hot, processed material.  

Such a wind may be driven by some of the white dwarfs themselves
(Fields, Freese, and Graff \cite {ffg2}).
Some of the white dwarfs may accrete from binary red giant companions
and give rise to Type I Supernovae, which serve as an energy
source for Galactic winds.  However, excess heavy elements
such as Fe may be produced in the process (Ruiz--Lapuente \cite{pilar}).

\subsection{On Carbon and Nitrogen}
\label{sect:carbon}
The issue of carbon (Gibson \& Mould \cite{gm}) and/or nitrogen
produced by white dwarf progenitors is the
greatest difficulty faced by a white dwarf dark matter scenario.
Stellar carbon yields for zero 
metallicity stars are quite uncertain.
Still, according to the Van 
den Hoek \& Groenewegen (1997) yields, a star of mass 
2.5$\msun$ will produce about twice the 
solar enrichment of carbon.  
If a substantial fraction of all 
baryons pass through intermediate mass stars, the carbon abundance in this 
model will be near solar.

Then overproduction of carbon can be a serious problem,
as emphasized by Gibson \& Mould \pcite{gm}.
They noted that stars in our galactic halo have carbon 
abundance in the range $10^{-4}-10^{-2}$ solar, 
and argued that the gas 
which formed these stars cannot have been polluted by the ejecta of a 
large population of white dwarfs.
The galactic winds discussed in the previous section could
remove carbon from the star forming regions and mix it throughout the 
universe.  

However, carbon abundances in intermediate redshift 
\lya\ forest lines have recently been measured to be 
quite low.
Carbon is indeed present, but only at the
$\sim 10^{-2}$ solar level,
(Songaila \& Cowie \cite{sc}) for \lya\ systems at $z \sim 3$
with column densities $N \ge 3 \times 10^{15} \, {\rm cm}^{-2}$.
Furthermore, in 
an ensemble average of systems 
within the redshift interval $2.2 \le z \le 3.6$,
with lower column densities 
($10^{13.5} \, {\rm cm}^{-2} \le N \le 10^{14} \, {\rm cm}^{-2}$),
the mean C/H drops to $\sim 10^{-3.5}$ solar
(Lu, Sargent, Barlow, \& Rauch \cite{lsbr}).

In order to maintain carbon abundances as low as $10^{-2}$ solar, only about 
$10^{-2}$ of all baryons can have passed through the intermediate mass 
stars that were the predecessors of Machos.  Such a fraction can barely
be accommodated by our results in section 4.1  
for the remnant density predicted from our extrapolation 
of the Macho group results, and would be in conflict with
$\Omega_{{\rm prog}}$ in the case of a single burst of star formation.

We note that progenitor stars lighter than 4$\msun$ overproduce Carbon;
whereas progenitor stars heavier than 4$\msun$ may replace the carbon
overproduction problem with nitrogen overproduction (Fields, Freese,
and Graff \cite{ffg2}).   The heavier
stars may have a process known as Hot Bottom Burning, in which
the temperature at the bottom of the star's convective envelope
is high enough for nucleosynthesis to take place, and carbon
is processed to nitrogen (Lattanzio \cite{latt},
Renzini and Voli \cite {rv}, Van den Hoek and Groenewegen (1997),
Lattanzio and Boothroyd \cite{lattbooth}). 
In this case one gets a ten times
solar enrichment of nitrogen, which is far in excess of the
the observed nitrogen in damped Lyman systems.
In conclusion, both C and N exceed what's obersved.

Note that it is possible (although not likely) that carbon never leaves
the white dwarf progenitors, 
so that carbon overproduction is not a problem
(Chabrier, private communication).
Carbon is produced exclusively in the stellar core.  
In order to be ejected, carbon must convect to the outer layers in 
the ``dredge up'' process.  Since convection is less efficient in a zero 
metallicity star, it is possible that no carbon would be ejected in a 
primordial star.  In that case, it would be impossible to place limits on 
the density of white dwarfs using carbon abundances. 
We have here assumed that carbon does leave the white dwarf progenitor
stars.

\subsection{Neutron Stars}
The first issue raised by neutron star Macho candidates
is their compatibility with the microlensing results.
Neutron stars ($\sim 1.5 \msun$) and
stellar black holes ($\ga 1.5 \msun$)
are more massive objects, so that one would typically expect longer
lensing timescales than what is currently observed in
the microlensing experiments (best fit to $\sim 0.5 \msun$).
As discussed by Venkatesan, Olinto, \& Truran \pcite{vot},
one must posit that
as the experiments continue to take measurements, longer
timescale events should begin to be seen.
In this regard, it is intriguing that the
first SMC results (Palanque-Delabrouille et al.\ \cite{eros:smc};
Alcock et al.\ \cite{macho:smc}) suggest lensing masses
of order $\sim 2 \msun$.  

However, the same issues of mass budget and chemical
enrichment arise for neutron stars as did
for white dwarfs, only the problems are worse.
In particular, the higher mass progenitors of neutron
stars eject even more mass, so that
$\Omega_{\rm prog}$ is even bigger than for the case of white dwarfs.
The ejecta are highly metal rich and would need a great
deal of dilution (as much as for the case of white dwarfs)
in order to avoid conflict with observations. However,
most of the baryons in the universe have already been used to 
make the progenitors (even more than for the
case of white dwarfs); there are no baryons left over to do the
diluting.

\subsection{Mass Budget Summary:}

If Machos are indeed found in halos of galaxies like our own,
we have found that the cosmological mass budget for Machos
requires $\omegam/ \omegab \geq {1 \over 6} h f_{\rm gal}$,
where $f_{\rm gal}$ is the fraction of galaxies that contain Machos,
and quite possibly $\omegam \approx \omegab$.  
Specifically, the central values in
eqs.\ (\ref{omega}) and (\ref{omegab}) give
$\omegam/\omegab \sim 0.7$.
Thus a stellar explanation of the
microlensing events requires that a significant
fraction of baryons cycled through Machos and their
progenitors. If the Machos are
white dwarfs that arose from a single burst of star
formation, we have found that the contribution of the progenitors
to the mass density of the universe is at least a factor of two higher,
probably more like three or four.  
We have made a comparison of $\omegab$ with
the combined baryonic component of $\omegam$  
and the baryons in the \lya\
forest, and  found that the values
can be compatible only for the extreme values of the parameters.
However, measurements of $\Omega_{{\rm Ly} \alpha}$
are at present uncertain, so that it is perhaps premature to
imply that Machos are at odds with the amount of baryons in 
the Ly$\alpha$ forest.
In addition, we have stressed the difficulty in reconciling the Macho mass
budget with the accompanying carbon and/or nitrogen
production in the case of white
dwarfs.  The overproduction of carbon or nitrogen
by the white dwarf progenitors
can be diluted in principle, but this dilution would require
even more baryons that have not gone into stars.  
At least in the simplest scenario, 
in order not to conflict with the upper bounds on $\omegab$,
this would require an $\omegam$ slightly smaller
than our lower limits from extrapolating the Macho results.
Only 10$^{-2}$ of all baryons can have passed through
the white dwarf progenitors, a fraction that is in conflict
with our results for $\Omega_{{\rm prog}}$.

\section{Zero Macho Halo?}
The possibility exists that the 14 microlensing events
that have been interpreted as being in the Halo of the Galaxy
are in fact due to some other lensing population.  One of the most
difficult aspects of microlensing is the degeneracy
of the interpretation of the data, so that it is
currently impossible to 
determine whether the lenses lie in the Galactic Halo, or in the Disk
of the Milky Way, or in the LMC.  Evans \etal (\cite{evans:diskflare}) proposed
that the events could be due to lenses in our own Milky Way Disk.
Gould (\cite{gouldy}) showed that the standard model of the LMC does
not allow for significant microlensing.  

Zhao (1998) has proposed that debris lying in a tidal tail stripped
from the progenitor of the LMC or SMC by the Milky Way or by an
SMC-LMC tidal interaction may
explain the observed microlensing rate towards the LMC. Within this
general framework, he suggests that the debris thrown off by the
tidal interaction could also lead to a high optical depth for
the LMC.  There have been several observational attempts to search for
this debris.  Zaritsky \& Lin (1997) report a possible detection of
such debris in observations of red clump stars, but the results of
further variable star searches by the {\sc macho} group (Alcock \etal
1997b), and examination of the surface brightness contours of the LMC
(Gould 1998) showed that there is no evidence for such a population.
A stellar evolutionary explanation for the observations of Zaritsky \&
Lin (1997) was proposed by Beaulieu \& Sackett (1998).  However,
possible evidence for debris within a few kpc of the LMC along the
line of sight is reported by the {\sc eros} group (Graff \etal in
preparation).  These issues are currently unclear and are under
investigation by many groups.

Note that a recent microlensing event towards the SMC, MACHO-98-SMC-1,
was due to a binary lens.  In this case it was possible to clearly
identify that the lens is in the SMC  and not in our Halo (Albrow 
\cite{planet}).
So far, there are 2 SMC events, both in the SMC.  However, the
situation for sources in the LMC remains ambiguous and awaits
further observations.

\section{Conclusions}
Microlensing experiments have ruled out a large class of 
possible baryonic dark matter components.  As described
in the contribution by Mark Moniez, substellar objects
in the mass range $10^{-7}\msun$ all the way up to $10^{-2}\msun$
are ruled out by the experiments.  In this talk
I discussed the heavier possibilities in the range $10^{-2}\msun$
to a few $\msun$.  I showed that brown dwarfs and faint stars
are ruled out as significant dark matter components; they contribute
no more than 1\% of the Halo mass density.  White dwarfs and
neutron stars still survive as a possible explanation for the Macho
events that have been seen, but only if one pushes the allowed
ranges of all the parameters.  

Hence, in conclusion,
\hfill\break
1) Nonbaryonic dark matter in our Galaxy seems to be required, and
\hfill\break
2) The nature of the Machos seen in microlensing experiments
and interpreted as the dark matter in the Halo of our Galaxy remains a mystery.
Are we driven to primordial black holes (Carr 1994), nonbaryonic
Machos (Machismos?), or perhaps a no-Macho Halo?

\bigskip
We are grateful for the
hospitality of the Aspen Center for Physics,
where part of this work was done.
DG acknowledges the financial support
of the French Ministry of Foreign Affairs' Bourse Chateaubriand.
KF acknowledges support from the DOE at the
University of Michigan.  
The work of BDF was
supported in part by
DOE grant DE-FG02-94ER-40823.

Figure Captions:

\medskip

Figure 1 (taken from Dahn et al 1995):
H-R diagram of nearby stars with measured parallax.
The filled circles are high metallicity disk stars;
the velocity dispersion of these disk stars is 
$\sim 30$ km/sec.
The open circles are low metallicity halo ``subdwarfs";
these stars have high proper motions $\sim 200$ km/sec.
We have superimposed a solid line which indicates
the theoretical model of Baraffe et al (1995) with log(Z/Z$_\odot ) = -1.5$.

\medskip

Figure 2 (taken from Graff and Freese 1996a):
Distances to six stars in HST data with $1.7 < V-I < 3$
obtained by comparing apparent with absolute magnitudes
of these stars.  

\medskip

Figure 3: (taken from Graff and Freese 1996b):
The mass function of red dwarf halo stars (multiplied
by m$^2$).  Each of the four
models is derived from the LF of Dahn ${\it et \,
al}$ (1995) but assumes different metallicity and binary content.  In
all three panels, crosses without errorbars illustrate the mass
function derived for stars with metallicity Z = $3 \times 10^{-4}$ and
no binary companions.  
The other model presented in panel (a) has Z =
$6 \times 10^{-4}$ (no binaries) for comparison.  
Panels (b) and (c)
show binary models II and III for $Z = 3 \times 10^{-4}$.
Binary model III has
been designed to exaggerate the number of low mass stars compared to
high mass ones and is unrealistic.

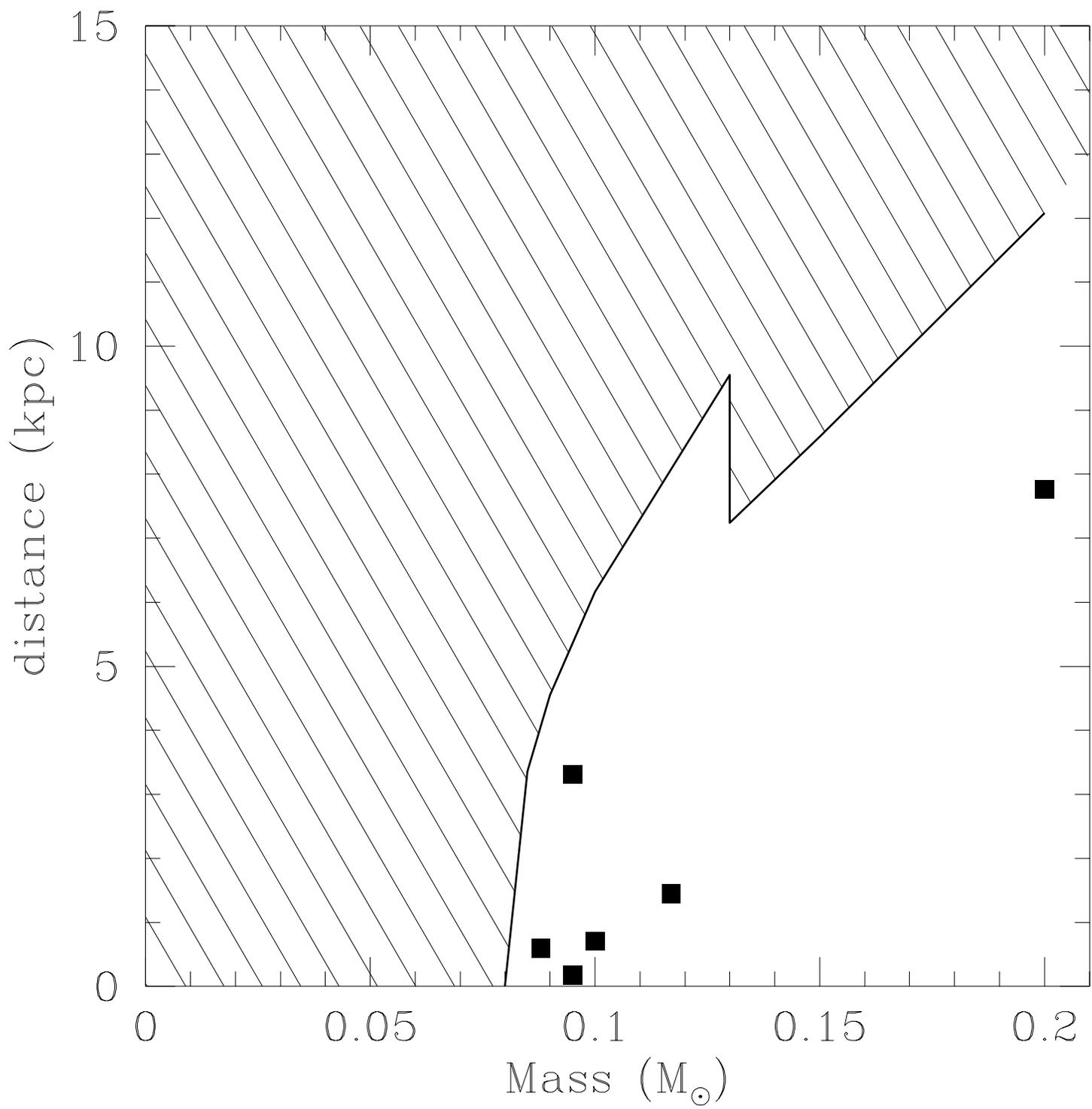
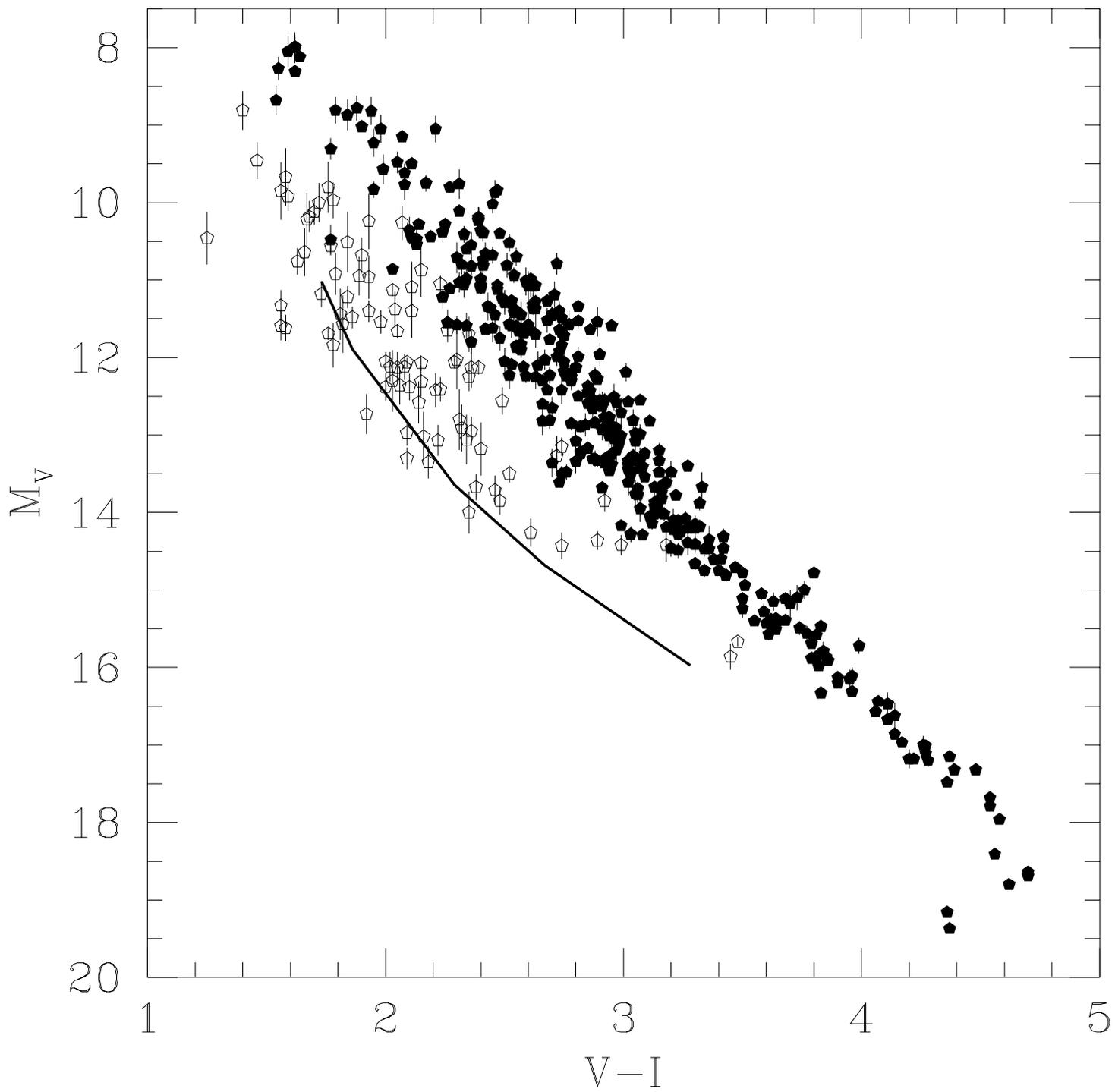
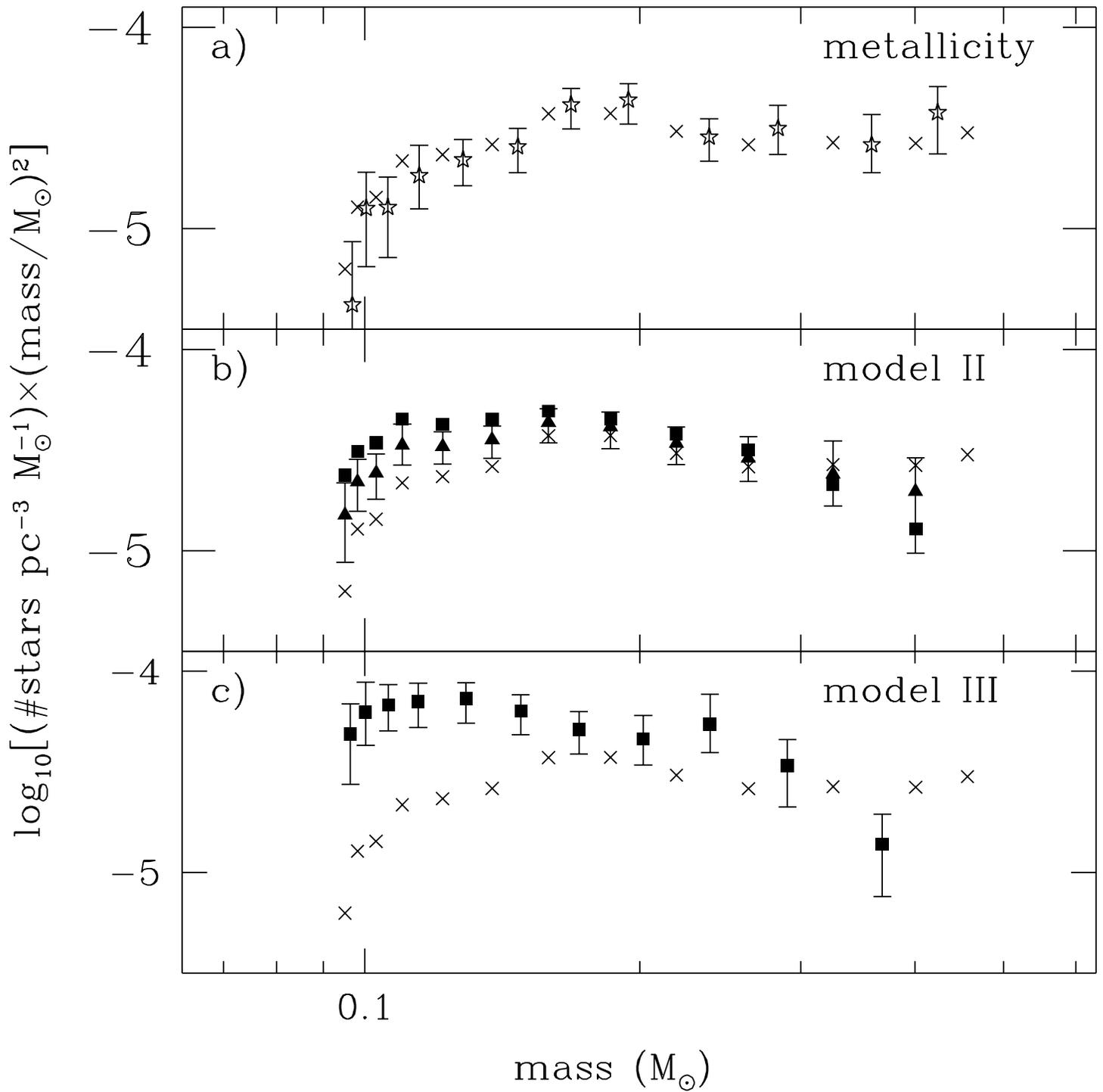
\end{document}